\documentclass[
 ,final            
  ]
  {aipproc}

\layoutstyle{8x11double}

\begin{document}

\title{$\beta$~Cephei Stars from ASAS: a New Look at Hot Pulsators}


\author{Andrzej Pigulski}{
  address={Instytut Astronomiczny Uniwersytetu Wroc{\l}awskiego, Kopernika 11, 51-622 Wroc{\l}aw, Poland}
}
\author{Grzegorz Pojma\'nski}{
  address={Obserwatorium Astronomiczne Uniwersytetu Warszawskiego, Al.~Ujazdowskie 4, 00-478 Warszawa, Poland}
}

\begin{abstract}
The All Sky Automated Survey (ASAS) data, covering nearly 70\% of the 
whole sky south of declination $+$28$^{\rm o}$, were recently used to 
discover about 300 $\beta$~Cephei stars brighter than V$\sim$11.5 mag.
As this means a nearly fourfold increase in the number of known stars of this 
type, new possibilities in studying these pulsators open up, including 
statistical work.

The homogeneity of the ASAS survey allows us to study their distribution 
in the local Galaxy.  We discuss this distribution in the context of the 
location of nearby spiral arms.  In addition, we show pulsational (periods, 
amplitudes) properties of the whole sample of known $\beta$~Cephei stars.  
Some objects interesting from the point of view of asteroseismology are also indicated.
\end{abstract}

\maketitle

\section{Introduction}
The most recent review paper on $\beta$~Cephei stars, young, massive and hot 
main-sequence pulsators was presented several years ago by \cite{StHa05}.  
They provided a list of 93 members of this group 
that were discovered during a century-long history of their investigation.  
The review was followed by several comprehensive multi-site campaigns focused on selected 
$\beta$~Cephei stars, namely those that had the largest number of modes known: 
$\nu$~Eri, 12 (DD) Lac, V836\,Cen, and $\theta$~Oph. These observations allowed to make 
a reliable mode identification and consequently enabled seismic modeling.

In the last few years, the first satellite observations of $\beta$~Cephei stars 
were also conducted. Several bright stars of this type ($\beta$~Cru, $\lambda$~Sco, 
$\kappa$~Sco, $\alpha$~Vir, $\beta$~CMa, $\alpha$~Lup, $\nu$~Eri, $\beta$~Cep) 
were observed by the WIRE satellite tracker \cite{Brun07}.  One of the most 
interesting results of these observations was the confirmation of the eclipsing 
nature of $\lambda$~Sco \cite{BrBu06}. Next, the Canadian MOST satellite observed
several $\beta$~Cephei stars: $\delta$~Cet \cite{Aert06,Jerz07,Mold09}, $\gamma$~Peg
\cite{Hand09} and Spica which was found to be eclipsing \cite{Desm09}. Finally, 
GSC\,06272-01557 = ALS\,4680 was discovered as a new $\beta$~Cephei star \cite{Anto09}. 
It was independently found in the ASAS data (see the next section), where only two strongest modes were
detected. The third satellite that had $\beta$~Cephei stars in the target list was
CoRoT. It observed V1449~Aql \cite{Degr09}, the only $\beta$~Cephei star known in the CoRoT
fields at the time of target selection.

During the last few years, some serendipitous discoveries as well as regular searches,
especially in the young open clusters, increased the number of known $\beta$~Cephei stars by additional 19 
objects. However, the largest impact on the number of known $\beta$~Cephei stars came from
the All Sky Automated Survey (ASAS).  In this paper, we show how this survey contributed 
to our knowledge of this group of pulsating stars.

\section{$\beta$~Cephei stars from ASAS}
The ongoing, third part of the ASAS, ASAS-3 \cite{Pojm01,Pojm09}, covers about 70 per cent of
the sky south of declination $+$28$^{\rm o}$.  It started in 2000 and already 
brought the discovery of over 50\,000 variable stars published in the form of the catalog 
\cite{Pojm02,Pojm03,PoMa04,PoMa05,Pojm05}. The stars in the catalog were classified by
means of an automatic classification scheme that utilized mainly periods, 
amplitudes and near-infrared photometry of variable stars.
The $\beta$~Cephei as a group was first included in the third paper of the series 
\cite{PoMa04}.  However, since $\beta$~Cephei stars are low-amplitude pulsators, the 
automatic classification was not always sufficient to distinguish
unambiguously a $\beta$~Cephei star from the other low-amplitude short-period variable stars.  
Thus, this classification was verified by \cite{Pigu05} who used additional information
on the ASAS variables spread over the literature, mainly spectral types and/or $UBV$
photometry.  In consequence, 14 $\beta$~Cephei stars were found. In addition, \cite{Hand05}
did the same for the ASAS-2 data \cite{Pojm00} finding five more stars of this type.

All these nineteen $\beta$~Cephei stars found in the public ASAS data had relatively 
large amplitudes, namely larger than $\sim$~30~mmag.  This was the consequence of the method
of the selection of objects for the ASAS variable-star catalogue, namely by means of the magnitude-dispersion diagram.  
On the other hand, the quality and number of the ASAS data allowed detection of periodic 
signals with amplitudes down to a few mmag. An obvious conclusion was that many more 
$\beta$~Cephei stars could be discovered using the ASAS data.  The search for these 
variables was done in two steps.  First, about 15\,000 O and early B-type stars were selected using
over two hundred catalogs with suitable information (mainly on spectral types and multi-colour 
photometry).  Next, the data for the selected stars were analyzed in detail.  The result 
largely exceeded the expectations: 276 $\beta$~Cephei stars were discovered. The first part of 
this work was already published: \cite{PiPo08} presented the list of 103 new $\beta$~Cephei stars.
A detailed description of the remaining stars is in the course of preparation for publication. 

In the following subsections we present the main properties of the whole group of 406 $\beta$~Cephei
stars that are currently known, dividing them on two parts: those discovered with the use
of the ASAS data (295 stars) and the remaining 112 stars. For clarity, we will refer to these
two groups as ASAS and non-ASAS $\beta$~Cephei stars, respectively.

\subsection{Pulsational properties}
\begin{figure}
  \includegraphics[width=3.1in]{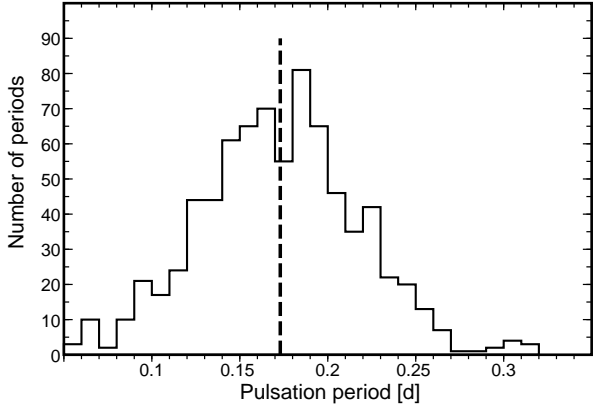}
  \caption{Period distribution for modes detected in 406 known $\beta$~Cephei stars. The vertical 
   dashed line shows the mean and median value of the distribution equal to 0.173~d.}
  \label{p-distr}
\end{figure}

The most important pulsational characteristics of $\beta$ Ce\-phei stars include periods and amplitudes.
In the whole group of over 400 $\beta$~Cephei stars, we know over 750 independent pulsation modes. 
Distribution of their periods is shown in Fig.~\ref{p-distr}.  The periods longer than 0.35~d were
excluded from this figure. They are observed in hybrid $\beta$~Cephei/SPB pulsators and represent
most likely $g$ modes.  The distribution of periods is quite symmetric around the mean value of
0.173~d (shown with vertical dashed line in Fig.~\ref{p-distr}). The median value of the distribution is 
the same as the mean.

\begin{figure}
  \includegraphics[width=3.1in]{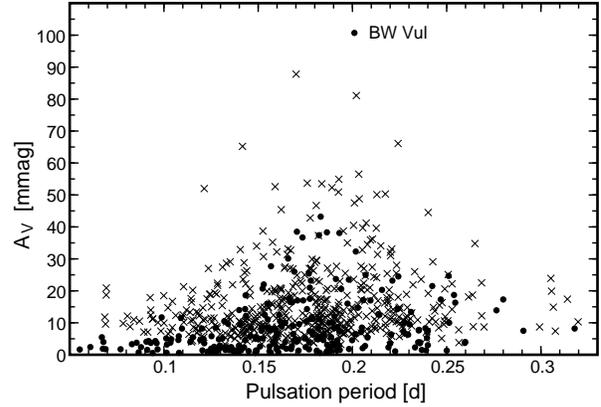}
  \caption{$V$-filter semi-amplitudes, $A_V$, for the same modes that are shown in Fig.~\ref{p-distr}
   plotted against their pulsation periods. Amplitudes for the ASAS $\beta$~Cephei stars are shown as
   crosses, for the remaining stars, as dots.  The star with the largest amplitude known, BW~Vul, is labeled.}
  \label{a-distr}
\end{figure}

The distribution of $V$-filter semi-amplitudes for the same sample of modes is plotted 
against the period of the mode in Fig.~\ref{a-distr}. In this figure, the data for ASAS and non-ASAS 
$\beta$~Cephei stars are shown with different symbols.  It can be clearly seen that most of the
large-amplitude pulsators were found in the ASAS data, although the well-known variable BW~Vul (labeled)
remained the star with the largest amplitude known. Nevertheless, the gap between 40 and 100~mmag has
been filled up with the modes detected in the ASAS $\beta$~Cephei stars.  With so many modes we may also safely conclude that 
0.1~mag is an observational upper limit for the $V$-filter semi-amplitude of a $\beta$~Cephei star.  Another 
feature that can be seen from Fig.~\ref{a-distr} is that stars with the largest amplitudes are found
in the middle of the range of periods observed in $\beta$~Cephei stars.  Close to the limits of this
range, the amplitudes tend to be smaller.  This can only be explained in view of the non-linear theory
of pulsations and the results of the first theoretical works \cite{SmMo07} seem to agree with the observed
distribution.

From the point of view of asteroseismology, the best are those stars that have both large amplitudes 
(mode identification is easier for them) and many modes excited. It is even better if some modes are rotationally split
which can be recognized by the presence of multiplets equidistant or almost equidistant in frequency.
\cite{PiPo08} presented a list of seven stars with likely rotational splittings. In the second group of
analyzed stars, two more stars with almost equidistant triplets, ALS\,2932 and ALS 9911, were found.

\begin{figure}
\caption{Distribution of the ASAS (crosses) and non-ASAS $\beta$~Cephei stars (filled circles) in Galactic coordinates shown in Aitoff projection.
The size of the circle depends on the magnitude of a star: brighter stars are shown with larger circles.
The Galactic center is located in the middle of the oval. The sky covered by the ASAS observations ($\delta < +$28$^{\rm o}$) is shown in gray and
delimited by a dashed line. Solid line is the celestial equator.}
\includegraphics[width=6.4in]{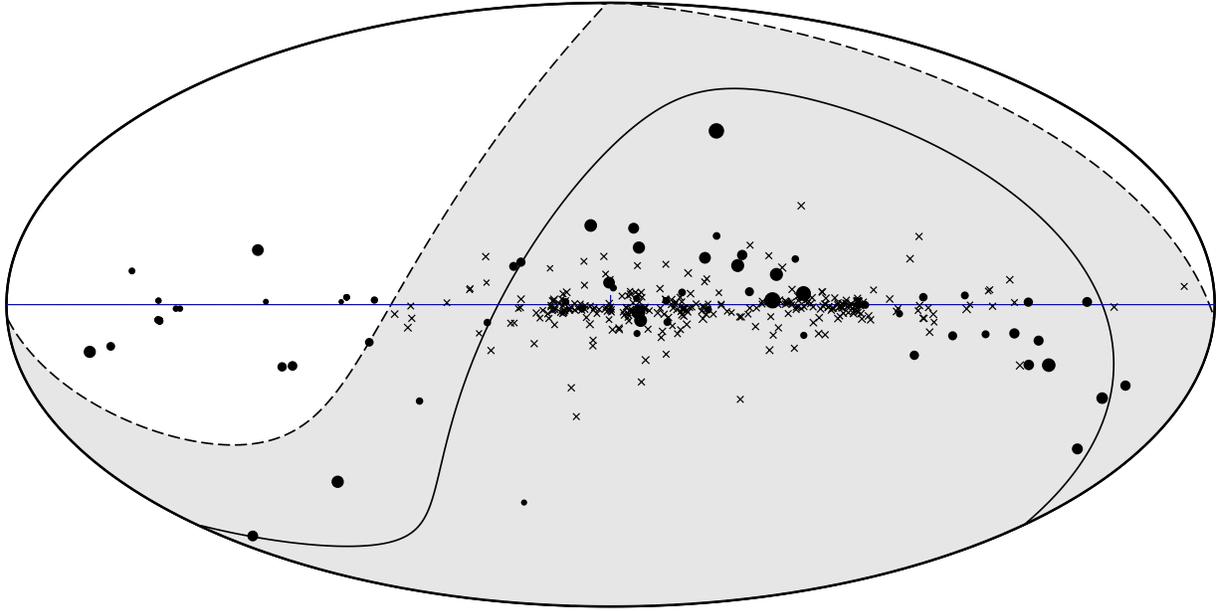}
\label{sky-distr}
\end{figure}

Another group of stars that can be very useful in seismic modeling are hybrid $\beta$~Cephei/SPB stars, i.e.~stars
that exhibit both $p$ (high-frequency) modes intrinsic to $\beta$~Cephei stars and $g$ (low-frequency)
modes characteristic for the SPB stars.  Several candidates for hybrid $\beta$~Cephei/SPB stars among the ASAS
$\beta$~Cephei stars were indicated by \cite{PiPo08}.  Again, some more were found in the second group.  One of them,
HDE\,313073, is especially interesting as it shows two modes in the low-frequency (0.4--0.5~d$^{-1}$) 
and two modes in the high-frequency (7.2--7.4~d$^{-1}$) domain. It is interesting to note that the
ASAS candidates for hybrids have, at least for the low-frequency modes, much larger amplitudes than those found
for such modes in the bright hybrids 12 Lac, $\nu$~Eri, and $\gamma$~Peg \cite{Hand09}. 

\begin{figure}
\caption{The ASAS $\beta$~Cephei stars (dots) projected onto the Galactic plane. The Sun is located in the center of the arcs
that cover the range of galactic longitudes corresponding to the part of the sky observed by the ASAS. The location of
spiral arms was adopted following \cite{Chur09}; they were labeled with common names. The short, unlabeled arm 
is the Orion Spur.}
\includegraphics[width=3.1in]{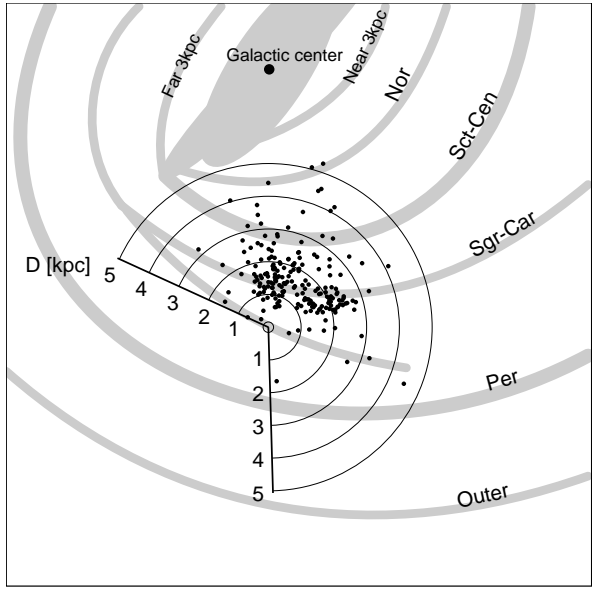}
\label{gplane-distr}
\end{figure}

\subsection{Distribution in the sky and the Galaxy}
Distribution of the ASAS and non-ASAS $\beta$~Cephei stars in the sky is shown in Fig.~\ref{sky-distr}. Since 
$\beta$~Cephei stars are young objects, their concentration towards the Galactic plane is obvious. On the 
other hand, the ASAS $\beta$~Cephei stars do not spread uniformly along the Galactic plane.  They are located mainly at galactic
longitudes, $l$, between 285$^{\rm o}$ and 20$^{\rm o}$ with a lower density for 310$^{\rm o}$\,$< l <$\,330$^{\rm o}$. 
The clumping is partially due to the non-uniform extinction, but the main effect is a consequence of the young 
age of these stars.  Since young stars are distributed largely close to the star-forming regions, they trace the location
of spiral arms.  For this reason, so few $\beta$~Cephei stars were discovered for $l <$\,285$^{\rm o}$, where there is
no nearby spiral arm.  This can be better seen in Fig.~\ref{gplane-distr}, where 
the ASAS $\beta$~Cephei stars were projected onto the Galactic plane. The distances of stars were estimated using the $UBV$
photometry in the same way as in \cite{Pigu05}.  As can be seen, most of the stars have distances in the range of 1--3 kpc
and they are located in the Sagittarius-Carina spiral arm.  However, stars up to the distance of 5 kpc are observed. 
Some of them might be runaway stars, which are located at high Galactic latitudes as can be seen in Fig.~\ref{sky-distr}.
 
\section{Conclusions}
The ASAS provided a large number of bright $\beta$~Cephei stars and many of them are good targets for asteroseismology. 
Their $V$ magnitudes range between 8 and 12 mag. This is a very
convenient range for the follow-up photometric and spectroscopic work. They are neither too bright for photometry
with a small telescope nor too faint for spectroscopy with a moderate-size one.  In addition, many of them have 
large amplitudes and are multiperiodic.  The problem is that the ASAS $\beta$~Cephei stars 
are mostly southern objects and a very limited number of sites with the appropriate telescopes is currently available. 
The discovery, however, might be a very good reason for preserving small telescopes in the southern sites and devote them 
to the follow-up photometry and/or spectroscopy of $\beta$~Cephei stars.  

The ASAS survey was recently extended to the
Northern hemisphere \cite{Pojm09}. Thus, a similar analysis will be soon possible for the whole northern sky.  In particular, we
may expect discovering a large number of $\beta$~Cephei stars located in the Perseus arm (see Fig.~\ref{gplane-distr}).  It must be noted,
however, that due to the poor spatial resolution of the ASAS frames, the ASAS photometry in very dense regions is not very good.
Consequently, the ASAS $\beta$~Cephei stars are mostly field objects or objects located in sparse open clusters and associations.  
The searches for these stars in young open clusters, some ongoing (see, e.g.~\cite{Kola04} and references therein),
can be therefore stated as a good complementary work.

\begin{theacknowledgments}
This work was supported by the grant N\,N203\,302635 from Polish MNiSzW.
\end{theacknowledgments}

\bibliographystyle{aipprocl} 

\end{document}